\documentclass[nofootinbib,aps,letterpaper,superscriptaddress,showpacs]{revtex4}
\usepackage{amsmath}
\usepackage{amsfonts}
\usepackage{amssymb}
\usepackage[dvips]{graphicx}

\begin{document}

\newcommand\be{\begin{equation}}
\newcommand\ee{\end{equation}}
\newcommand\ba{\begin{eqnarray}}
\newcommand\ea{\end{eqnarray}}
\newcommand\bseq{\begin{subequations}} 
\newcommand\eseq{\end{subequations}}
\newcommand\bcas{\begin{cases}}
\newcommand\ecas{\end{cases}}
\newcommand{\p}{\partial}
\newcommand{\f}{\frac}
\newcommand{\nn}{\nonumber \\}
\def\tr{{\rm Tr}}

\title{Triangulated Loop Quantum Cosmology:\\ Bianchi IX and inhomogenous perturbations}

\author{Marco Valerio Battisti}
\email{battisti@icra.it}
\affiliation{Centre de Physique Th\'eorique, Case 907 Luminy, 13288 Marseille, EU}
\author{Antonino Marcian\`o}
\email{antonino.marciano@cpt.univ-mrs.fr}
\affiliation{Centre de Physique Th\'eorique, Case 907 Luminy, 13288 Marseille, EU}
\author{Carlo Rovelli}
\email{rovelli@cpt.univ-mrs.fr}
\affiliation{Centre de Physique Th\'eorique, Case 907 Luminy, 13288 Marseille, EU}

\begin{abstract}
\noindent 
We develop the ``triangulated" version of loop quantum cosmology, recently introduced in the literature. We focus on the ``dipole" cosmology, where space is a three-sphere and the triangulation is formed by two tetrahedra. We show that the discrete fiducial connection has a simple and appealing geometrical interpretation and we correct the ansatz on the relation between the model variables and the Friedmann-Robertson-Walker scale factor. The modified ansatz leads to the convergence of the Hamiltonian constraint to the continuum one.  We then ask which degrees of freedom are captured by this model.  We show that the model is rich enough to describe the (anisotropic) Bianchi IX Universe, and give the explicit relation between the Bianchi IX variables and the variables of the model. We discuss the possibility of using this path in order to define the quantization of the Bianchi IX Universe. The model contains more degrees of freedom than Bianchi IX, and therefore captures some inhomogeneous degrees of freedom as well.  Inhomogeneous degrees of freedom can be expanded in representations of the $SU(2)$ Bianchi IX isometry group, and the dipole model captures the lowest integer representation of these, connected to hyper-spherical harmonic of angular momentum $j=1$. 
\end{abstract}

\maketitle

\section{Introduction}

Loop quantum cosmology (LQC) \cite{revLQC} is the most remarkable application of the loop approach to quantum gravity. Its main result is a robust indication that the cosmological singularity 
that appears in classical general relativity is removed by quantum effects  \cite{Ash_Big_Bounce}.  
While the physical basis of this results is explicitly grounded on the physical discreteness of quantum geometry which is predicted  \cite{LAV} by full loop quantum gravity (LQG) \cite{Rovelli_book}, the precise relation between LQC and LQG, on the other hand, has not been fully clarified yet \cite{Bo1}. 
To shed light on this issue, a finite dimensional truncation of LQG, which leads naturally to LQC in the Born-Oppenheimer approximation, was introduced in \cite{Rovelli&Vidotto1}. The idea is to fix a coarse triangulation of physical space and consider a discretization and a quantization of general relativity on this triangulation. This procedure leads to a truncated version of LQG, interpreted as a description of a finite number of large-scale (namely cosmological) degrees of freedom. In the simplest version of the theory, called the ``dipole'' cosmology, compact physical space is triangulated with two tetrahedra. It was shown in \cite{Rovelli&Vidotto1} that this model leads to a LQC-like dynamics in the Born-Oppenheimer approximation, where the scale factor plays the role of the ``heavy" degree of freedom.  The discrete dynamics of LQC is recovered in this manner without recurring to the ``area gap'' argument. 

The main purpose of this work is to analyze this dipole cosmology more in detail, and in particular, interpret the geometrical meaning of its other (``light'') degrees of freedom. It has been suggested that these could just capture some space anisotropies. 
We show that the model captures indeed all the anisotropic degrees of freedom, that is, the degrees of freedom of Bianchi IX.  This may be of particular interest since Bianchi IX describes a {\it generic} space-time near the cosmological singularity \cite{BKL}.  We write an explicit relation between the  Bianchi IX degrees of freedom and those of the model. We discuss the possibility of using this relation in order to construct a loop quantization of the Bianchi IX cosmological model. 

However, the model captures a richer dynamics than Bianchi IX: it includes inhomogeneous degrees of freedom as well, and therefore truly represents a ``first step towards inhomogeneity'' in LQC.  These degrees of freedom can be identified as the lowest order term of a tensor harmonic expansion, in terms of Wigner functions, of the spatial geometry (see Ref.~\cite{Bojw}). The physical relevance of this expansion is  of interest not only in cosmology, but also in relation with the study of the LQG $n$-points correlation functions, which is based on a similar approximation \cite{distretisation-quantization,propagator}. 

In the course of our analysis, we obtain also two other results.  First, we show that the discrete version of the fiducial connection used to define the theory, which was defined in a rather \emph{ad-hoc} manner in \cite{Rovelli&Vidotto1}, has in fact a neat and appealing geometrical interpretation on the triangulated space.  Second, following \cite{bodava}, we correct the naive Bohr-Oppenheimer ansatz used in \cite{Rovelli&Vidotto1} to relate the model variables to the scale factor. We show that using a different ansatz the Friedman equation can be obtained explicitly in the appropriate limit, from the discrete Hamiltonian constraint.  Therefore the model agrees with the standard formulation of LQC both at the classical and quantum levels. The triangulated Hamiltonian constraints can be interpreted as a different version of the LQC constraint.  

We recall that although most of the LQC literature focuses on the isotropic sector \cite{Bo3, Asht_FRW_k=1}, homogeneous \cite{Bo4, AAWE, AAWE2} and inhomogeneous \cite{Bo5, Gowdy_applications} models have been studied as well.

The paper is organized as follows.  In Section \ref{Bianchi9} we review the geometry of the Bianchi IX Universe. In Section \ref{model} we review the dipole cosmology model.  In Section \ref{interpraetatio} we relate the the dipole cosmology with Bianchi IX.  In Section \ref{FRW} we study the classical limit of the model and show that it reproduces the classical theory at small curvature. In Section \ref{anis}, we study the way anisotropies are described in the model. Finally, inhomogeneous degrees of freedom are introduced in Section \ref{inhomo}. 
 
We set $8 \pi G/3=1=c$, where $G$ is the Newton gravitational constant and $c$ the speed of light.  We use small Latin letters $a,b,c,...$ as spatial indices on the three-dimensional surfaces, and capital Latin letters $I,J,K...$ to label internal indices of $su(2)$-algebra-elements, which are raised and lowered with the identity matrix. Sum over internal or external indices is meant whenever those are repeated, even if they are not paired,  and do not appear inside brackets. When indices are within brackets, summation is \emph{not} be intended. That is $a_Ib^Ic^I\equiv\sum_Ia_Ib^Ic^I$, but $a_{(I)}b^{(I)}c^{(I)}=d^I$. We consider Euclidean gravity and we set for simplicity the Barbero-Immirzi parameter as $\beta=1$.

\section{Bianchi IX in a nutshell} \label{Bianchi9}

In this Section we recall the basic properties of the Bianchi IX cosmological model, which plays a central role in the following. For a detailed discussion about the topology of this model we refer to the Appendix \ref{tresfera}. 

The Bianchi IX model is the most general homogeneous model for a spatially compact Universe.
It is also the basis of the 
classical description of a generic inhomogeneous space-time near the singularity, via the Belinski-Khalatnikov-Lifshitz (BKL) scenario \cite{BKL}:  
as the cosmological singularity is approached spatial points (causal horizons) decouple dynamically and each of them evolves essentially independently as a Bianchi IX model \cite{BKL}. Thus classical physics is well described in terms of this model near the cosmological singularity.

Consider a four-dimensional metric space-time $\mathcal{M}$ and let $\phi: \mathcal{M} \to \mathbb{R} \otimes \Sigma_t$ be a diffeomorphism, where $\Sigma_t$ are the Cauchy surfaces foliating space-time. This is defined to be spatially homogeneous if for any $t\in  \mathbb{R}$ and any two points $p,\,q \in \Sigma_t$ there exists an isometry of the space-time metric which takes $p$ into $q$. 
A Bianchi IX cosmology is a homogeneous cosmology where  $\Sigma_t$ has the topology of the three sphere  $S^3$. 

In order to write homogeneous fields on $S^3$, it is convenient to have a homogeneous reference triad field.  This can be constructed by exploiting the fact that $S^3$ can be identified with the group manifold of $SU(2)$.  This manifold carries the natural Cartan (flat) connection 
\be \label{sciaca}
\omega=g^{-1}dg=\omega^{I} \tau_I=\omega_a^{I} \tau_I dx^a\,,  
\ee
where $g\in SU(2)$ and $x^a$ are three arbitrary coordinates on $SU(2)$. Here $\tau_I=\sigma_I/(2i)$ is a basis in the $su(2)$ Lie algebra and $\sigma_I$  are the Pauli matrices. This connection is left-invariant and satisfies the Maurer-Cartan structure equation
\be \label{MC}
d\omega^{I}  - \frac{1}{2} \epsilon^{\!I} _{\,\,\,JK} \, \omega^{J} \wedge \omega^{K} =0\,,
\ee 
where $\epsilon_{IJK}$ is the completely antisymmetric tensor. 
Denote  $e_{I}=e^a_{I}\partial_a$ the corresponding dual vector field, with values in $su(2)$, such that $e^a_{I}\omega^{J}_a=\delta_{I}^{J}$. The Lie brackets of these vector fields read
\be
[e_{I},\,e_{J} ]= -\epsilon_{\,\,\,IJ}^{K} \,e_{K}.
\ee
By identifying  $su(2)$ with an internal space, we can take $e_{I}$ and $\omega^I$ as the definition of a triad and co-triad in space, which are invariant under the (left) action of $SU(2)$ on itself. These fields can be used as ``fiducial" co-frames and frames, as they naturally carry information about the homogeneity symmetry group.  

An explicit parametrisation for the co-frames can be given in terms of Euler angles as
\ba \label{omega}
\omega_1&=&\cos \psi \,d \theta + \sin \psi \sin \theta\, d\phi\,,\\ \nonumber
\omega_2&=&\sin \psi \,d\theta - \cos \psi \sin \theta \, d \phi\,,\\  \nonumber
\omega_3&=& d \psi + \cos \theta\, d \phi\,.
\ea
Here $\theta$ and $\phi$ are in the range $\theta \in [0, \pi)$, $\phi \in [0, 2 \pi)$, while for the Euler angle $\psi$ one must set $\psi\in [0, 4 \pi)$ in order to achieve a simply connected covering space with the topology of $S^3$.
This parametrization allows us to make easily contact with the definition of Cartan connections in terms of elements $g\in SU(2)$. These latter can in fact written as
\ba \label{gigi}
g &=& \begin{pmatrix}\alpha & -\beta\\
              \beta^\star& \alpha^\star
            \end{pmatrix}\,,  \qquad   {\rm with}     \nonumber\\  
\alpha&=& e^{\f{i}{2}(\phi+\psi)}\, \cos (\theta/2)\,, \qquad
\beta= e^{-\f{i}{2} (\phi-\psi)}\, \sin (\theta/2)\,,
\ea
from which the above expression for the co-frames follows by means of (\ref{sciaca}). One can also easily find the expression for the left invariant frame $e^a_{I}$ in this chart, as well as the right invariant vector fields (the Killing vector fields) which Lie drag the above  frame and co-frame.

Any homogeneous Riemannian metric $q_{ab}$ on the Cauchy surfaces can be then written in terms of its triadic projection $q_{IJ}$ on internal space as
\be
 q_{ab}(x,t)=q_{IJ}(t) \, \omega^{I}_a(x)\,\omega^{J}_b(x),
\ee
where $q_{IJ}$ is a $3\times3$ matrix, constant in space. 
In the isotropic case $q_{IJ}$ is a multiple of the identity, and it is proportional to the Killing-Cartan metric on $SU(2)$.  
The space-time line-element can then be written in the form (taking the lapse function $N=1$)
\be \label{tuche}
ds^2= -dt^2+ q_{IJ} \, \omega^{I}\otimes\omega^{J}\,.
\ee
The vacuum Einstein equations allow us to write $q_{IJ}(t)$ in diagonal form \cite{Mix, MTW}
\be\label{metricbix}
q_{ab}= a^2_{I} \,\omega^{I}_a\omega^{I}_b,
\ee
where $a_{I}=a_{I}(t)$ are the three (independent) scale factors which describe the anisotropy of the space slices $\Sigma_t$. Thus homogeneity reduces the phase-space of general relativity to six dimensions. The inverse metric reads $q^{ab} =   a_I^{-2} \, e_{I}^a e_{I}^b $ and the spatial volume $\mathcal{V}$ is given in terms of the scale factors by $ \mathcal{V}= 2 \pi^2 a_{1}\,a_{2}\,a_{3}$. The closed Friedmann-Robertson-Walker (FRW) cosmological model is recovered as the particular case $a_1=a_2=a_3$.

The Ashtekar-Barbero variables for the Bianchi IX cosmological model are hence written in terms of the Maurer-Cartan connection (\ref{MC}) and the dual triad, and are parametrized by three time-functions $c^I(t)$ and their duals $p^I(t)$ as (see for instance \cite{Bo4,bodava})
\ba \label{AE}
A^I_a=c_{(I)}\, \omega^{(I)}_a\,, \qquad E^a_I=p^{(I)}\,\omega\, e^a_{(I)}\,,
\ea
in which enters the determinant of the co-triad $\omega\equiv\det(\omega^I_a)$, namely the square root of the determinant of the metric of the three sphere. The connections $c_I$ and momenta $p^I$ parametrize the six-dimensional phase space, the symplectic two-form being $\Omega=dc_I\wedge dp^I$. The momenta are related to the metric variables by the relations
\be
p_1=|a_2a_3|\text{sgn}(a_1), \qquad p_2=|a_1a_3|\text{sgn}(a_2), \qquad p_3=|a_1a_2|\text{sgn}(a_3)\,,
\ee
and the connections are given in terms of triadic projection of Christoffel symbol components $\Gamma_I$ and extrinsic curvature components $K_I=-\dot a_I/2$ as $c_I=\Gamma_I-K_I$. The $\Gamma_I$ are here given in terms of the scale factors $a_I$ as 
\be
\Gamma_I=\f12\left(\f{a_J}{a_K}+\f{a_K}{a_J}-\f{a_I^2}{a_Ja_K}\right)=\f12\left(\f{p^K}{p^J}+\f{p^J}{p^K}-\f{p^Jp^K}{(p^I)^2}\right).
\ee  
In the closed FRW model \cite{Asht_FRW_k=1} ($a_1=a_2=a_3$) the triadic projection of the Christoffel symbols becomes the constant $ \Gamma = 1/2$ and one recovers the isotropic connection $c=(\dot a+1)/2$ as well as the momentum $|p|=a^2$. Thus the cosmological singularity in Bianchi IX appears whenever $a_I=0$ for some $I$.

\section{A model for merging LQC in LQG} \label{model}

We briefly recall the construction of the model \cite{Rovelli&Vidotto1}. 

\subsection{Classical theory}

Fix an oriented triangulation $\Delta_n$ of the topological three-sphere, formed by $n$ tetrahedra $t$ glued by their triangles. Label the triangles with an index  $f$ (``$f$" for \emph{face}) that runs from 1 to $2n$ (the number of faces is twice the number of tetrahedra).  A group element $U_f\in SU(2)$ and a $su(2)$ algebra element $E_f$ are associated to each oriented triangle $f$. Given the face $f^{-1}$ obtained inverting the orientation of $f$, we take the convection that its associated group and algebra elements read
\be \label{inverso}
U_{f^{-1}}=U^{-1}_{f}\,, \qquad E_{f^{-1}}= - \ U_{f}^{-1}E_{f}U_{f}.
\ee
We use the notation $E_f=E_f^I\tau_I$ and take $U_f$ and $E_f$ as the phase space variables of a dynamical system. The phase space of this model is that of a canonical lattice $SU(2)$ Yang-Mills theory, {\it i.e.} the fundamental Poisson brackets are given by
\ba \label{simplec}
\{  U_f,U_{f'} \} = 0,  \qquad
\{  E_f^I,U_{f'} \} =  \delta_{\!f\!f'} \ \tau^I U_{f},  \qquad 
\{  E_f^I,E_{f'}^J \} = - \  \delta_{\!f\!f'}\  \epsilon^{IJK} E_f^K.  \qquad    
\ea
In other words, the phase space is the cotangent bundle of $SU(2)^{2n}$ with its natural symplectic structure.  The dynamics of the system is defined by two sets of constraints. The Gau\ss\, (gauge) constraint (three constraints per tetrahedron)
\be\label{gauge}
  G_t \equiv  \sum_{f\in t} E_f \approx 0\,,
\ee
where the sum is over the four faces of the tetrahedron, and the Hamiltonian constraint
\be\label{clham}
C_t \equiv  V^{-1}_t \sum_{ff'\in t} \tr[U_{\!f\!f'}E_{f'}E_{f}] \approx 0\,.
\ee
In (\ref{clham}) the sum is over the couples of distinct faces at each tetrahedron, $U_{\!f\!f'}=U_{f}U_{f'}^{-1}$ and  $V_t^2 =  \tr[E_fE_{f'}E_{f'\!'}]$ because of (\ref{gauge}). Here $V_t$ can be interpreted as (proportional to) the volume of the tetrahedron $t$ --- thus we call $V=\sum_t V_t$ the total volume of space. The Hamiltonian constraint gives the good classical limit once the gauge constraint has been taken into account. In fact, at the first order the holonomy is  $U\sim \exp\int_{\gamma}A\sim\, 1\!\! 1 - |\gamma|^2 F +O(|\gamma|^4 A^2)$, where $F$ is the curvature of the connection $A$ and $\gamma$ denotes a loop. By means of the holonomy expansion,  equation (\ref{clham}) can be formally recast as follows 
\be\label{hamf}
V_t C_t=\sum_{ff'\in t} \tr[E_{f'}E_{f}] -|\gamma|^2 \sum_{ff'\in t} \tr[F_{ff'}\,E_{f'}E_{f}] \approx 0.
\ee
The former term in relation (\ref{hamf}) vanishes because of the Gau\ss\, constraint, {\it i.e.} $\sum_{ff'\in t} \tr[E_{f'}E_{f}]= \tr[(\sum_{f\in t}E_{f})(\sum_{f'\in t}E_{f'})] \approx 0$, and thus the second term undergoes the expected continuum limit. We stress that this happens not only for small values of the length of the loop $|\gamma|$, but it does for large values of $|\gamma|$ too, provided that $|\gamma|^2 F$ is small.

This model can be regarded as a lattice approximation of the geometrodynamics of a closed Universe. To see this, consider real Ashtekar connection $A^I_a$ and electric field $E^a_I$, with their standard Poisson algebra (see for instance \cite{Rovelli_book}), on a three-dimensional surface $\Sigma$ with the $S^3$ topology.  Let $\Delta_n$ be a triangulation of  $\Sigma$ and $\Delta^*_n$ a dual of the triangulation. Then, $U_f$ is the parallel transport of the Ashtekar-Barbero connection along the link $e_f$ of $\Delta^*_n$ dual to the triangle $f$. In the same way, $E_f$ is the flux $\Phi_f$ of the conjugate electric field across the triangle $f$ (parallel transported  to the center of the tetrahedron). The Poisson brackets of $U_f$ and $E_f$ are thus exactly the ones in (\ref{simplec}). For instance, the third equality in (\ref{simplec}) is motivated by the parallel transport of $\Phi_f$ to the center of each tetrahedron. Of course these variables transform via the gauge constraint (\ref{gauge}). 

The constraint (\ref{clham}) corresponds to the non-graph-changing version of the Hamiltonian constraint. Its quantization is an operator acting on an underlying spin-network state representing the graph dual to the triangulation taken in consideration on the spatial manifold. Whenever the triangulation is sufficiently fine, constraint (\ref{clham}) represents an approximation the to Euclidean part of the Hamiltonian constraint $\tr[F_{ab}E^aE^b]/\sqrt{\det E}\approx 0$.

\subsection{Quantum theory}

The quantization of the model is straightforward.  A quantum representation of the observable algebra (\ref{simplec}) is provided in the auxiliary Hilbert space $\mathcal{H}_{aux}=L_2[SU(2)^{2n}, dU_f]$, where $dU_f$ is the Haar measure.  That is, states have the form $\psi(U_f)$.  The operators $U_f$ are diagonal and the operators $E_f$ are given by the left invariant vector fields on each $SU(2)$ element. Consistently, operators $E_{f^{-1}}$ act as right invariant vector fields. Not surprisingly, the operator associated to the volume $V_t$ turns out to be the standard LQG volume operator \cite{Rovelli_book}. 

States which are solutions of the Gau\ss\, constraint (\ref{gauge}) are labeled by SU(2) spin networks on the graph $\Delta_n^*$. The dual triangulation $\Delta_n^*$ is characterized by a node for each tetrahedron and a link for each face of the triangulation $\Delta_n$. 
In the dual triangulation only four-valent vertices appear, which intertwine spin-$j_f$ representations $D_{j_f}(U)$ associated to the holonomies around the links $l_{f}$. The basis of these spin-network states is labeled by $|j_f, \iota_t\rangle$, where $\iota_t$ denotes the intertwiner quantum number at the given node. The spin network basis states are explicitly given by
\be\label{spinnet}
\psi_{j_{\!f}\iota_t}\!(U_f)=\langle U_f | j_f, \iota_t \rangle=\otimes_f\  D^{(j_f)}(U_f)  \cdot  \otimes_t \ \iota_t\,, 
\ee
in which ``$\cdot$" indicates the contraction of the indices of the $D^{(j_f)}(U)$ matrices with the indices of the intertwiners $\iota_t$.

This constrained model can be directly quantized \emph{\`a la} Dirac. The Hamiltonian constraint  can be defined \emph{\`a la} Thiemann, rewriting (\ref{clham}) in the form 
\be\label{hamth}
C_t  =  \  \sum_{ff'f'\!'\in t} \epsilon^{f\!f'\!f'\!'}\ \tr[U_{\!f\!f'}U^{-1}_{f'\!'}\{U_{f'\!'},V_t\}] \approx 0
\ee 
and then defining the corresponding quantum operator by replacing the Poisson bracket with the commutator.  Otherwise, there exists a second possibility: to implement directly as a quantum operator the constraint (\ref{clham}) rescaled by the volume $V_t$ 
\be\label{ham}
\tilde C_t = V_t C_t = \sum_{ff'\in t} \tr[U_{\!f\!f'}E_{f'}E_{f}] \approx 0\,. 
\ee
This is exactly the early proposal to perform the quantization of the Hamiltonian constraint in LQG, see \cite{smolincr}.
Physical states are those annihilated by the quantum version of (\ref{ham}), {\it i.e.} we have to impose $ \tilde C_t \psi = 0$. 
The Hamiltonian constraint on the whole space $\Sigma$ is obtained combining the set of $n$ Hamiltonian constraints by means of the lapse function $N=\{N_t\}$ at each node. Namely, there is $\tilde{C}(N) \psi \equiv \sum_t N_t \tilde{C}_t\,\psi=0$, $\forall\, N$. 

\section{Dipole cosmology} \label{interpraetatio}

A topological three-sphere can be constructed by gluing together the boundaries of two  three-balls (more details in Appendix \ref{tresfera}). The boundary of a three-ball is a two-sphere, and these two two-spheres are identified, with opposite orientations.  The boundaries of the two balls define then an ``equator" of the three-sphere. Accordingly, a three-sphere can be triangulated by gluing together two tetrahedra, along all their faces. The equator of the three-sphere is then triangulated by four triangles $f$, where $f=1,2,3,4$. (This is a cellular complex decomposition \cite{AlgeTopo}, and not a $\Delta$-complex or a simplicial complex decomposition \cite{Ruth, Coxter}.)
The dual graph of this triangulation described above is formed by two nodes joined by four links 
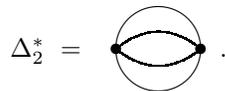
\begin{center}
\begin{picture}(20,40)
\put(-36,17) {$\Delta_2^*\  =$}
\put(20,20) {\circle{30}}
\put(04,20) {\circle*{4}} 
\put(36,20) {\circle*{4}}  
\qbezier(4,20)(20,33)(36,20)
\qbezier(4,20)(20,7)(36,20)
\end{picture} \hspace{2em}\raisebox{17pt}{.}
\end{center} 
From now we focus on the theory defined by this triangulation. 

The unconstrained phase space of the theory defined by this triangulation has twenty four dimensions and is coordinatized by $(U_f, E^I_f)$. At each node there are one Hamiltonian constraint $C\approx 0$ and three Gau\ss\, ones $\mathcal{G}^I\approx0$. But it is easy to verify that the constraints of the two nodes are in fact the same, giving a total of four constraints only. This brings the number of degrees of freedom down to eight. 
The Poisson bracket algebra between the Hamiltonian constraints closes in this case, as the Hamiltonian constraints at each node are actually the same one because of (\ref{inverso}). 

The Hilbert space of the quantum theory is $L_2[SU(2)^4/SU(2)^2]$. The spin network states that solve the gauge constraint are given by states $|j_f, \iota_t\rangle=|j_1,j_2,j_3,j_4,\  \iota_1, \iota_2\rangle$.  The action of the single Hamiltonian constraint gives in general $\tilde C |j_f, \iota_t\rangle = \sum_{ff'} C_{ff'}|j_f, \iota_t\rangle $. The Hamiltonian operator acts at each node.  The action of the Hamiltonian quantum constraint operator implies that
\be
C_{12}|j_1,j_2,j_3,j_4, \iota_1, \iota_2\rangle 
=\sum_{\epsilon,\delta=\pm 1}
C^{\epsilon\delta \iota'_1\iota'_2 }_{j_f\iota_1\iota_2} \ |j_1+\frac\epsilon{2},j_2+\frac\delta{2},j_3,j_4, \iota'_1, \iota'_2\rangle\,, 
\ee
in which matrix elements $C^{\epsilon\delta \iota'_1\iota'_2 }_{j_f\iota_1\iota_2} $ can be computed using recoupling theory. In terms of the wave function components, a more compact notation is given by 
\be
\tilde C\, \psi(j_f, \iota_t)\  
=\sum_{\epsilon_j=0,\pm 1}
{C}^{\epsilon_f \iota_t'}_{j_f \iota_t}\  \psi\!\left(j_f+\frac{\epsilon_f}{2},\iota'_t\right)\,.
\ee
Notice that $C^{\epsilon_j \iota_t'}_{j_f \iota_t}$ vanishes unless $\epsilon_f=0$ for only two $j$.   Matter fields can be simply added in this picture (see \cite{Rovelli&Vidotto1}).

\label{background}

In \cite{Rovelli&Vidotto1}, a discrete fiducial algebra element $\omega_f^I$ associated to the triangulation was introduced, and used to compare the variables of the dipole model with the FRW variables.  Here we introduce a novel and more useful definition for $\omega_f^I$. 
Consider the Plebanski two-form of the connection $\omega$
\be\label{veceta}
\Sigma^I(\omega)=\frac{1}{2}\epsilon^I\,_{JK}\,\omega^J \wedge \omega^K= \frac{1}{2}\,\omega\, e^{aI}\epsilon_{abc}\ dx^b\wedge dx^c\,,
\ee
where $\omega=\det(\omega_a^I)$, 
and let $\omega_f^I$ be the surface integral of this two-form on the triangle $f$ of the triangulation.  Using the Maurer-Cartan equation (\ref{MC}), we have
\ba \label{flatus}
\omega_{f}^I \equiv  \int_f \Sigma^I = \frac{1}{2}\,\int_f  \epsilon^I\,_{JK}\,\omega^J \wedge \omega^K = 
\int_f  d \omega^I = \oint_{\partial f}  \omega^I. 
\ea
That is, the flux of the Plebanski two-form across a triangle is equal to the line integral of $\omega^I$ along the boundary of the triangle. An immediate consequence of this is that 
for each tetrahedron $t$
\be
\sum_{f\in t} \omega_{f}^I =0\,,
\label{sum}
\ee
where the sum is over the triangles that bound the tetrahedron $t$. This is 
because the boundary of a boundary vanishes.  The set of $su(2)$ vectors $\omega_f^I$ form a natural background fiducial structure for the discrete theory, analogous to the $\omega^I$ fiducial connection in the continuous theory. Notice that integral of $|\Sigma|=\sqrt{\Sigma^I\, \Sigma_I}$  on a triangle $f$ is (twice) the area of $f$ determined by the background triad. 

In the case of the dipole, the explicit properties of the $\omega_f^I$ can be found using the symmetries. The action of $SU(2)$ that transforms the triangles into one another preserves the equator of the three-sphere.  The equator of $SU(2)$ is formed by the $\pi$ rotations around a direction $\vec n$. This set of rotations are transformed into one another by the adjoint action of the group on itself (which rotates $\vec n$ preserving the rotation angle). A discrete subgroup of this adjoint action sends therefore the triangles into one another. Under this adjoint action, $\omega^I$ (which is in the algebra) transforms under the adjoint representation, which is the fundamental representation of $SO(3)$. That is, the four vectors $\omega_f^I$ are rotated into each other by rotations (of the $I$ index) of the same angles. Therefore they are proportional to the normals of a regular tetrahedron in $\mathbb R^3$. The cosine of the angle between two such vectors is therefore $1/3$.  It is also convenient to take $\omega^I$ so that the norm of $\omega_f^I$ is one: $\omega^I_f\,\omega^I_f\equiv |\omega_f|^2=1$.  This characterizes entirely the $\omega_f^I$, up to an overall rotation.

\section{Isotropy: closed FRW model} \label{FRW}

We now begin the analysis of the physics and geometry of the model.  We start, in this section, with the restriction of the model to the homogeneous and isotropic sector. This analysis was already presented in \cite{Rovelli&Vidotto1}, but we show here that it can be strongly ameliorated by using a different ansatz on the relation between the model variables and the scale factor.  Anisotropies and inhomogeneities are discussed in the following sections. 

\subsection{Classical framework}

Let $c(t)$ and $p(t)$ denote the isotropic connection variables used in LQC \cite{revLQC, Asht_FRW_k=1}. They are related to the scale factor $a$ and to its time derivative $\dot a$ by $c=(\dot a+1)/2$ and $|p|=a^2$ (see Section \ref{Bianchi9}). The phase space of an homogeneous and isotropic cosmology is two-dimensional and the basic Poisson bracket is $\{c,p\,\}=1$. 

In \cite{Rovelli&Vidotto1}, the isotropic phase space variables $(c,p)$ are identified with a subspace of the dipole model phase space by the relations
\be  \label{ttt}
U_f=\exp\left(c\,\omega^I_f\tau_I\right), \qquad E_f=p\,\omega^I_f\tau_I.
\ee
These relations define an embedding of the isotropic geometries in the
phase space of the model.  Notice that this is different from what is
done in \cite{AAWE2}, where a projection of the
anisotropic degrees of freedom of the Bianchi I model to those ones of the FRW model was defined.
(See also \cite{NelSak} for a discussion about superselection
of isotropic FRW states of LQC within the anisotropic Bianchi I
setting.)

The second of the relations (\ref{ttt}) follows immediatly from \eqref{AE}, \eqref{flatus} and the identification of $E_f$ with the flux of the electric field $E$ though $f$ (Section \ref{model}). The discussion of the first relation is more delicate. Recall that $U_f$ is the holonomy of the connection along a dual link $e_f$, 
\be \label{Uu} 
U_f=\mathcal{P} \exp \int_{e_f}\,A^I_a(e_f) \tau_I\, de_f^a\, .
\ee
Let us assume for simplicity that the identity in the group $G$ is identified with the center of one of the two tetrahedra. Then the four dual links $e_f$ run along one-parameter subgroups of $G$. 
Now consider a point $g$ on the $SU(2)$ group manifold. Because of the flatness of the Maurer-Cartan connection $\omega$, the holonomy of $\omega$ from the identity to the point $x_P$ does not depend on the particular path which has been chosen to connect the identity with $g$.  We can write $g=\exp ( \alpha^I \tau^I)$, in which $\alpha^I=\alpha^I(g)$ is a vector in the internal space, and then parametrize the path $\gamma$ from the identity of the group to $g$ by 
\be \label{ggg}
g(s)=\exp \left( s\, \alpha^I \tau^I  \right)\,,
\ee
in which $s\in[0,1]$. This particular path defines an abelian subgroup of $SU(2)$, and thus the holonomy reads
\be
U_f\equiv \mathcal{P} \exp \int_{1\!\!1 \rightarrow{g(s)}}\!\!\!\!\! \omega^I \tau^I=\,\,g(s)\,.
\ee
That is: the holonomy of $\omega$ from the identity to $g$ is precisely $g$. Consider now the tangent to the path $\gamma$ at $g$, and denote it $v^a\in T_g$ ($T_g$ is the tangent space of the group).  Since we have a metric structure  $\!\!\!\phantom{a}^{o}q_{ab}=\omega^I_a\omega^I_b$ defined on the group manifold, we have also the corresponding one-form $n_a=\!\!\!\phantom{a}^{o}q_{ab}\,v^b$. 
The contraction of $\omega$ and $v^a ds$  is given by
\be
\omega(v^a\,ds)=  n_a e^{a}_{I} \tau^I \, ds\,.
\ee
On the other hand, by means of (\ref{sciaca}) and (\ref{ggg}), we find
\be
\omega(v^a\,ds)= g^{-1} dg(v^a\, ds)= \alpha^I\tau^I ds\,.
\ee
Hence
\be e^{a}_{I} n_a= \alpha^I. 
\ee 
Now, take $g$ to be the intersection between the dual link $e_f$ and the triangle $f$. By symmetry, the link $e_f$ and the triangle $f$ are orthogonal.  The last equation shows that $\alpha^I(g)$ is nothing else than the the normal $n_a$ to the triangle $f$, with the index raised by the fiducial triad.  But $\omega_f$ is precisely the integral of $e^{a}_{I} n_a$ on the triangle, and we expect it to be proportional to the value of $e^{a}_{I} n_a$ on $g$. Therefore we can conclude that 
\be
U_f = \exp\left(k \omega^I_f\tau_I\right),
\ee
for some $k$. Following \cite{bodava}, differently from what discussed in \cite{Rovelli&Vidotto1} we make the ansatz
\be   \label{giusta}
U_f=\exp\left((c+\alpha)\,\omega^I_f\tau_I\right)\,,
\ee
where $\alpha$ is a numerical constant determined by the spatial curvature, which we fix below. We refer to \cite{bodava} for a discussion of the rational of this identification. 

The Gau\ss\, constraint reads
\be
\mathcal G^I=\sum_f E^I_f=\sum_f p\,\omega^I_f=0\,,
\ee
and is automatically satisfied because of (\ref{sum}).
The scalar constraint (\ref{clham}) reads
\be
\tilde C=p^2\sum_{ff'}\text{Tr}\left[e^{c\,\omega^I_f\tau_I}e^{\alpha\,\omega^J_f\tau_J}e^{-\alpha\,\omega^K_{f'}\tau_K}e^{-c\,\omega^L_{f'}\tau_L} \omega^S_{f'}\,\omega^M_f\,\tau_S\,\tau_M\right]\approx0. 
\ee
Using the relation $e^{c\,\omega^I_f\tau_I}=\cos(c/2)+2\omega^I_f\tau_I\sin(c/2)$ the expression above can be simplified. Firstly notice that the term containing two $\tau$-matrices is zero because of the Gau\ss\, constraint. The terms including the traces of three and five $\tau$-matrices are also zero because of the internal product of two equal vectors $\omega^I_f$ and of the Gau\ss\, constraint. Moreover the term proportional to the trace of six $\tau$-matrices identically vanishes. (See Appendix \ref{tau} for details about traces of $\tau$-matrices.) Finally, as already discussed, each $\omega_f^I$ identifies a normal to the center of the face $f$ of the tetrahedron, the angle between them given by $\cos\theta_{ff'}=\omega^I_f\omega^I_{f'}=1/3$. Collecting these considerations and using the relations
\be
\sum_{ff'}\omega^I_f\omega^I_{f'}=\sum_{ff'}\cos\theta_{ff'}=0, \qquad \sum_{ff'}(\omega^I_f\omega^I_{f'})(\omega^J_f\omega^J_{f'})=\sum_{ff'}\cos^2\theta_{ff'}=\f23, \qquad 
\sum_{ff'}|\omega_f|^2|\omega_{f'}|^2=6,
\ee
we obtain the final form of the scalar constraint
\be\label{confrw}
\tilde C=\f{17}6p^2\left(\cos(c-\alpha)-1\right)\approx0.
\ee
Such a constraint describes the dynamics of a (curved) triangulated closed FRW cosmological model. The ordinary classical dynamics can be recovered as soon as the limit of small connection $|c|\ll1$ is taken into account. The constraint (\ref{confrw}) rewrites as 
\be\label{classisocon}
\tilde C=\f{17}6p^2\left(c\sin\alpha-\f{c^2}2\cos\alpha-2\sin^2(\alpha/2)\right)+\mathcal O(p^2c^3)\approx 0.
\ee
This is the FRW classical constraint $\tilde C_{FRW}\propto- p^2c(c-1)$ as soon as $\alpha$ is fixed at a suitable value $\alpha^\star$. In fact, the canonical transformation $c\rightarrow ac+b, \, p\rightarrow p/a$ ($a,b\in\mathbb R$) with  $\cos\alpha^\star=(9-\sqrt{17})/8$ transforms the two constraint into one another. The appearance of the ordinary dynamics for small values of the connection $c$ is in agreement with the claim that a coarse triangulation well approximates the classical theory for a low-curvature space-time. 
 
The expression (\ref{confrw}) can be regarded as the effective Hamiltonian constraint of the standard LQC. The effective formulation of LQC is usually obtained by ``polymerizing'' the classical model: the connection is replaced by its exponentiated version, to reflect the fact that the connection operator does not exist in the LQC Hilbert space \cite{Corichi}. The effective LQC dynamics can be also obtained by using the methods of geometric quantum mechanics \cite{tav}. Replacing $c\rightarrow \sin(\nu c)/\nu$ (where $\nu$ denotes the polymer scale, fixed by the ``minimal area gap'' argument \cite{Corichi, AACoSi}) in $\tilde C_{FRW}$, one obtains the constraint (\ref{confrw}) for $\alpha=\alpha^\star$. Here, instead, the ``polymerization" is a consequence  of the existence of the triangulation, and a conventional quantization of the  truncated theory.  In the quantum theory, a WDW difference equation arises without adding an input from LQG by hand: the lattice model leads directly to a difference evolution equation, without recurring to the ``minimal area gap" argument.  

\subsection{Quantum framework}\label{quantum}

The variable $c$ multiplies the generator of a $U(1)$ subgroup of the compact group $SU(2)^4$. Therefore it is a periodic variable: $c\in[0,4\pi]$. The kinematic Hilbert space $\mathcal{H}_{iso}$ of the theory is thus $L^2(S^1, dc/4\pi)$ of square integrable functions on a circle. Eigenstates of $\hat p$ (from now on we drop the hat in order to simplify the notation) are labeled by an integer  $\mu$ and read $\langle c|\mu\rangle=e^{i\mu c/2}$. Wave functions $\psi(c)$ are decomposed in a Fourier series of eigenstates of $p$, labelled by an integer $\mu$
\be
\psi(c)= \sum_n\, \psi_\mu e^{i \mu c/2}\,.
\ee
The fundamental operators on this representation are  $p$ and $\exp(i c/2)$, whose action on generic states reads
\ba
p\, |\mu\rangle = \mu/2\ |\mu\rangle\,, \qquad  \exp(i c/2)\,|\mu\rangle=|\mu+1\rangle\,.
\ea
In particular, the operator $\sin(c/2)$ acts on $|\mu\rangle$ as
\be\label{sinop}
\sin(c/2)|\mu\rangle=\f1{2i}\left(|\mu+1\rangle-|\mu-1\rangle\right)
\ee
and it is then immediate to obtain the action of the composite operators $\sin(c/2)\cos(c/2)$ and $\sin^2(c/2)$. 

By following this path, we can construct the quantum constraint operator corresponding to (\ref{confrw}). In particular, in order for the states $\psi(c)$ to be physical states, the coefficients $\psi_\mu=\langle c|\,\mu\rangle$ have to satisfy the recurrence relation
\be\label{diffeq}
D^+(\mu)\,\langle c|\,\mu+2\rangle+D^0(\mu)\,\langle c|\,\mu\rangle +D^-(\mu)\,\langle c|\,\mu-2\rangle=0\,,
\ee
where 
\ba
D^+(\mu)&=&\f{17}6\mu^2\left(-i\sin(\alpha/2)\cos(\alpha/2)-\f12\sin^2(\alpha/2)+\f12\cos^2(\alpha/2)\right),\\\nonumber
D^0(\mu)&=&\f{17}3\mu^2\left(\f12-\sin^2(\alpha/2)\right),\\\nonumber
D^-(\mu)&=&\f{17}6\mu^2\left(i\sin(\alpha/2)\cos(\alpha/2)-\f12\sin^2(\alpha/2)+\f12\cos^2(\alpha/2)\right).
\ea
In (\ref{diffeq}) we have chosen the simplest normal ordering of posing multiplicative operators on the left of derivative ones. Equation (\ref{diffeq}) has the structure of the LQC difference equation. 

\section{Anisotropy: Bianchi IX model} \label{anis}

In this Section we free more degrees of freedom than the sole scale factor and we  describe the triangulated version of a homogeneous but non isotropic space-time. Since the space topology is $S^3$, we are therefore dealing with the Bianchi IX cosmological model. Bianchi IX (or the Mixmaster Universe \cite{Mix}) is the most general homogeneous model with this topology: its physical importance relies in describing a generic (classical) solution of the Einstein equations toward a space-like singularity via the BKL scenario \cite{BKL}. 

\subsection{Classical framework}

The dynamics of an anisotropic (homogeneous) cosmological model is described by three scale factors $a_I=a_I(t)$, which identify three independent directions (in the time evolution) of the Cauchy surfaces.  In the connection formalism, relaxing the isotropy condition corresponds to consider three different connections $c_I=c_I(t)$ and momenta $p_I=p_I(t)$. Our model can be then extended to an anisotropic setting by demanding that the variables of the theory are given by
\be\label{varbix}
U_f=\exp\left(c^I\omega^I_f\tau_I\right)\exp\left(\alpha\,\omega^I_f\tau_I\right), \qquad E_f=p^I\omega^I_f\tau_I.
\ee

Notice that there is a main difference between the holonomies (\ref{varbix}) and those used in LQC in the anisotropic context \cite{bodava, Bo4}. In the standard formulation of an anisotropic LQC model the basic holonomies $h_I$, are directional objects, computed along edges parallel to the three axis individuated by the anisotropies. They read 
\be
h_I=\exp(c^{(I)}\omega_{(I)})=\cos(c^I/2)+2\omega_I\sin(c^I/2).
\ee
On the other hand, the variables $U_f$ are non-directional objects, because the four faces of the triangolation do not have any special orientation with respect to the three isotropy axes. The connection components are summed over and they are thus independent on the $I$-direction. The $U_f$ are in fact group element of $SU(2)$ that depend on the face $f$. Explicitly, they are given by
\be \label{dava}
\exp\left(c^I\omega^I_f\tau_I\right)=\cos\left|\f{c^I\omega^I_f}2\right|+2\f{c^I\omega^I_f\tau_I}{|c^I\omega^I_f|}\sin\left|\f{c^I\omega^I_f}2\right|=\cos\rho_f+2\hat\rho^I_f\tau_I\sin\rho_f,
\ee 
where $\hat\rho^I_f$ denotes the unit vector $\hat\rho^I_f=\rho^I_f/\rho_f$ defined by
\be \label{diva}
\rho_f^{I}\equiv \frac{1}{2} c^{(I)} \omega^{(I)}_f\,, \qquad \rho_f=\sqrt{\sum_I (\rho_f^I)^2}=\f12|c^I\omega^I_f|.
\ee

Let us now analyse the dynamics of the anisotropic model. As above, the Gau\ss\, constraint does not carry out any information since it vanishes because of the construction of the vectors $\omega^I_f$. Explicitly, it is given by 
\be\label{gaussbix}
\mathcal G^I=\sum_f E^I_f=\sum_f p^{(I)}\omega_f^{(I)}=0.
\ee
The scalar constraint of the triangulated Bianchi IX model reads
\begin{multline}\label{conbix}
\tilde C=\sum_{ff'}\text{Tr}\left[(\cos\rho_f+2\hat\rho^I_f\tau_I\sin\rho_f)\left(\cos(\alpha/2)+2\omega^J_f\tau_J\sin(\alpha/2)\right)\right.\cdot\\ \cdot\left.\left(\cos(\alpha/2)-2\omega^K_{f'}\tau_K\sin(\alpha/2)\right)(\cos\rho_{f'}-2\hat\rho^L_{f'}\tau_L\sin\rho_{f'})
\xi^S_{f'}\xi^M_f\tau_S\tau_M\right]\approx0\,,
\end{multline}  
where the $\xi^I_f$ are defined by
\be\label{dyva}
\xi_f^{I}\equiv p^{(I)} \omega^{(I)}_f\,.
\ee
The difference with respect to the usual LQC constraint is evident. Let us note that, as a result of taking $U_f$ as defined above, no terms drop here in virtue of the Gau\ss\, constraint, as they did in the isotropic case. In fact, terms of the kind $g_fg_{f'}\omega_f\omega_{f'}$ (where $g_f$ are generic functions) appear and their sum in $f,f'$ no longer vanishes.
It is convenient to define
\be\label{angani}
\chi_{ff'}=\omega^I_f\hat\rho^I_{f'}, \qquad \beta_{ff'}=\omega^I_f\xi^I_{f'}, \qquad \gamma_{ff'}=\hat\rho^I_f\xi^I_{f'}, \qquad \sigma_{ff'}=\xi^I_f\xi^I_{f'}, \qquad \delta_{ff'}=\hat\rho^I_f\hat\rho^I_{f'}
\ee   
and
\ba\label{wedgevec}
A=\hat\rho\wedge\xi\wedge\xi, \qquad B=\omega\wedge\omega\wedge\xi, \qquad C=\omega\wedge\omega\wedge\hat\rho,\\\nonumber
\qquad D=\hat\rho\wedge\omega\wedge\xi, \qquad E=\hat\rho\wedge\hat\rho\wedge\omega, \qquad F=\omega\wedge\xi\wedge\xi.
\ea
where the convention is $A_{fff'}=\hat\rho_f\wedge\xi_f\wedge\xi_{f'}=\epsilon_{IJK}\hat\rho^I_f\xi^J_f\xi^K_{f'}$ and so on. The scalar constraint is given, in terms of all these quantities, by
\begin{multline}\label{conbixfin}
\tilde C=\sum_{ff'}\left\{\sin\rho_{f'}\cos\rho_f\left[A_{ff'f}\cos^2(\alpha/2)+\sin(\alpha/2)\cos(\alpha/2)\left(\beta_{ff'}\gamma_{f'f'}-\beta_{f'f'}\gamma_{ff'}+\f12\chi_{f'f'}\sigma_{ff'}+\beta_{ff'}\gamma_{ff'}+\right.\right.\right.\\
\left.\left.-\beta_{ff}\gamma_{f'f'}-\f12\chi_{ff'}\sigma_{ff'}\right)+2\sin^2(\alpha/2)\left(\f14\gamma_{ff}B_{ff'f}-\f14\gamma_{ff'}B_{ff'f'}+\sigma_{ff'}C_{ff'f'}+\cos\theta_{ff'}A_{f'f'f}\right)\right]+\\
+\cos\rho_f\cos\rho_{f'}\left[2\sin(\alpha/2)\cos(\alpha/2)F_{f'f'f}+\sin^2(\alpha/2)\left(\beta_{ff'}\beta_{ff'}-\beta_{ff}\beta_{f'f'}-\f12\cos\theta_{ff'}\sigma_{ff'}\right)-\f12\sigma_{ff'}\cos^2(\alpha/2)\right]+\\
+\sin\rho_f\cos\rho_{f'}\left[-\f12A_{ff'f}\cos^2(\alpha/2)+\sin(\alpha/2)\cos(\alpha/2)\left(\gamma_{ff'}\beta_{ff'}-\gamma_{ff}\beta_{f'f'}-\gamma_{ff'}\beta_{ff}+\gamma_{ff}\beta_{ff'}-\f12\chi_{ff'}\sigma_{ff'}+\right.\right.\\
+\left.\left.\f12\chi_{ff}\sigma_{ff'}\right)-\sin^2(\alpha/2)\left(\f14\beta_{f'f'}D_{fff}-\f14\beta_{ff'}D_{fff'}+\sigma_{ff'}C_{fff'}+\chi_{ff}F_{f'f'f}\right)\right]+\sin\rho_f\sin\rho_{f'}\left[\f{}{}\cos^2(\alpha/2)\cdot\right.\\
\cdot\left(\gamma_{ff'}\gamma_{ff'}-\gamma_{ff}\gamma_{f'f'}-\f12\delta_{ff'}\sigma_{ff'}\right)+\sin(\alpha/2)\cos(\alpha/2)\left(\f14\gamma_{f'f'}D_{ff'f}-\f14\gamma_{ff'}D_{ff'f'}-2\sigma_{ff'}E_{ff'f'}+\chi_{ff'}A_{f'f'f}+\right.\\
\left.+\f14\gamma_{ff'}D_{fff'}-\f14\gamma_{f'f'}D_{fff}-\chi_{ff}A_{f'f'f}\right)+\f12\sin^2(\alpha/2)\left(-\f12 E_{ff'f'}F_{ff'f}-\f12C_{fff'}A_{ff'f}+\sigma_{ff'}\left(\cos\theta_{ff'}\delta_{ff'}+\right.\right.\\
\left.\left.\left.\left.-\chi_{ff'}\chi_{ff'}\right)+\chi_{f'f'}\left(\gamma_{ff'}\beta_{ff}-\gamma_{ff}\beta_{ff'}\right)+\chi_{ff}\left(\gamma_{f'f'}\beta_{ff'}-\gamma_{ff'}\beta_{f'f'}\right)-4\chi_{ff}\chi_{f'f'}\sigma_{ff'}\f{}{}\right)\right]\right\}\approx0\,.
\end{multline} 
The dynamics of the Bianchi IX model, in the triangulation formalism, is then expressed in terms of geometrical quantities as angle-like (\ref{angani}) and volume-like (\ref{wedgevec}) quantities. As can be checked by direct calculation, the isotropic constraint (\ref{confrw}) is recovered as soon as $c_1=c_2=c_3$ and $p_1=p_1=p_3$. In this case all the terms in (\ref{wedgevec}) vanish because the internal wedge product of two equal elements $\omega_f$. Furthermore the terms containing any single angle (\ref{angani}) are also zero because of the Gau\ss\, constraint. 

As in the isotropic case, the constraint (\ref{conbixfin}) describes a discrete (triangulated) dynamics. Such a dynamics can be regarded as an effective one in which all the outcomes of the triangulation are abridged. The standard Bianchi IX model is recovered in the limit of small connections, {\it i.e.} as soon as $|c_I|\ll1$. In this limit, which corresponds to take $|\rho_f|\ll1$, the expression (\ref{conbixfin}) is rewritten as
\be\label{classlimbix}
\tilde C=p^Ip^J\left(const+c_K+c_Kc_L\right)+\mathcal O(p^2c^3)\approx0.
\ee
In order to make the notation the easiest as possible, constants (depending also on $\alpha$) are not written in (\ref{classlimbix}) and the contraction of the free indices is understood. The isotropic constraint (\ref{classisocon}) is recovered if a basic canonical transformation is performed, {\it i.e.} $c_I\rightarrow c_I+d_I$ and $p^I\rightarrow p^I+b^I$ for $d^I,b^I\in\mathbb R^3$. 

Summarizing, the constraint (\ref{conbixfin}) describes the effective LQG dynamics of an anisotropic homogeneous space-time having the spatial topology of a three-sphere, {\it i.e.} the Bianchi IX cosmological model. In the triangulation formalism the anisotropies (different scale factors or different connections) are described in terms of angles (\ref{angani}) and (internal) volumes (\ref{wedgevec}) characterizing a generic dynamical tetrahedron. The contact with the standard (classical) dynamics is recovered as soon as the flat space-time limit $|c_I|\ll1$ is taken into account.

\subsection{Quantum framework}

The fact that the dipole model includes Bianchi IX suggests that we may use it in order to construct a loop quantization of Bianchi IX.  We analyze this intriguing possibility in this section. There are two possible ways for doing so. One is to start from the $(c^I, p_I)$ variables, and define the dynamics with the Hamiltonian operator (\ref{conbixfin}), where all quantities are taken as functions of $(c^I, p_I)$ via equations (\ref{diva}) and (\ref{dyva}). The second possibility is to use the full quantum theory of the dipole cosmology model, defined in Section \ref{quantum}.  Neither of these approaches gives a simple quantum model, as far as we can see, since the resulting dynamical equation is complicated in both cases. 

Let us begin with a basis of states $|c^I\rangle$ that diagonalize the variables $c^I$, and write the quantum state as $\psi(c^I)$ in this basis. The $p^I$ operator is then $-i\partial/{\partial c^I}$.  For each face $f$, consider the three scaled variables (\ref{diva}) and the three scaled momenta (\ref{dyva}).  The Hamiltonian operator (\ref{conbixfin}) is entirely written in terms of the four operators 
$\xi_f^I$, $\sin\rho_f$, $\cos\rho_f$ and $\rho^I_f/\rho_f$.
These operators are easily written in this representation.  The first is 
\be
\xi_f^I= -i\omega_f^{(I)} \f{\partial}{\partial c^{(I)}}\,,
\ee
while the others are diagonal.  Thus, (\ref{conbixfin}) yields a Hamiltonian constraint on this Hilbert space, and defines a quantization of the Bianchi IX cosmological model. 

Does this quantization yield discreteness? Fix for a moment one of the four faces, say $f$. Observe that $\rho_f^I$ enters the dynamical theory only as an argument of the group element $U_f$ in (\ref{varbix}) via $U_f=\exp(2\rho_f^I\tau_I)\exp(\alpha\omega^I_f\tau_I)$. Therefore the physically relevant domain of $\rho_f^I$ appears to be given by 
\be
             |\rho_f^I|<4\pi
\ee
that is, adding $4\pi$ to $|\rho_f^I|$ gives back the same configuration of the $U_f$ variable (for a fixed $f$). Accordingly, the configuration space is compact, and we expect the conjugate variable $p_I$ to have discrete spectrum. This is reflected in the fact that the norm of  $\rho_f^I$ enters the Hamiltonian only via $\sin\rho_f$ (or $\cos\rho_f$). It seems therefore reasonable to expect that what is going on is something analogous to what happens in standard LQC, namely we can chose a scalar product where momenta have discrete spectrum, the operator $\sin\rho_f$ is well defined while $\rho_f$ is not, and the dynamics is restricted to a discrete set of eigenstates.  These observations are intriguing, but incomplete, because there are four distinct faces $f$, and four distinct group elements $U_f$, and generically, there is no shift on the $c^I$ that gives back the same value for \emph{all four} group elements. In other words, the map $c^I\to U_f$ defined in (\ref{varbix}) maps $\mathbb R^3$ into the compact space $[SU(2)]^4$, but its image is not necessarily compact. We leave the development of these reasonings for future investigations.

The second alternative is to use the dipole model Hilbert space $\mathcal H_d$, formed by states $\psi(U_f)$, and spanned by the discrete $|j_f,i\rangle$ basis. The reduction of a quantum theory to a sector where some degrees of freedom are frozen (here the inhomogeneous ones) is, in general, a rather non trivial step \cite{Bo1}.  If we have given the Bianchi IX Hilbert space $\mathcal H_{IX}$ formed by functions $\psi(c^I)$ and the  $c^I\to U_f$ map (\ref{varbix}), then a natural projection  $\pi:\mathcal H_d\to\mathcal H_{IX}$ is simply defined by $\psi(c_I)=\psi(U_f(c^I)).$ But in order to map $ \upsilon:\mathcal H_{IX}\to\mathcal H_d$ we need to choose a way to compute $c^I$ out of $U_f$.  One natural possibility is for instance to define 
\be
  c^I=\sum_f \ \frac{\rho_f^{(I)}}{2\omega_f^{(I)}}\,.
\ee
Using this, we can map any state in $\mathcal H_{IX}$ to $\mathcal H_d$. For instance, we can expand an eigenstate $\exp(i\mu^Ic^I)=\prod_I\langle c^I|\mu^I \rangle$ of $p^I$ on the spin network basis, considering $U_f=U_f(\rho^I_f)$ and formula (\ref{spinnet}), via
\be
\prod_I\,\langle j_f,i| \mu^I \rangle = \int_{SU(2)^4} dU_f
\left[\otimes_f \overline{D^{j_f}\left(U_f\right)\cdot i \cdot i}\,\right]\  \exp\left(i\sum_I\mu^I\sum_f \ \frac{\rho_f^{(I)}}{2\omega_f^{(I)}}\right).
\ee
Using these coefficients, the Bianchi IX Hamiltonian operator can be simply defined by
$\tilde C_{IX}=\pi \tilde C \upsilon$, where $\tilde C$ is the dipole Hamiltonian constraint operator, defined in the Section \ref{inhomo}.  This operator will be studied more in detail elsewhere.

\section{Inhomogeneity: perturbation of Bianchi IX} \label{inhomo}

Let us finally come to the full geometrical interpretation of the ``dipole model''. The remaining large-scale gravitational degrees of freedom captured by the dipole dynamics are necessarily  inhomogeneous. The easiest way to characterize them is to consider an expansion of the gravitational fields in tensor harmonics, and identify them with the lowest order term of this expansion. In other words, we can imagine that we want to describe only some low order term of a mode expansion of the geometry over the three sphere, and that the dipole variables read out these terms. In \cite{ReggeHu}, Regge and Hu considered a similar mode expansion around Bianchi IX, using Wigner $D$-functions. (See also \cite{tsh}.) Here we follow their approach, adapting it to the first order formalism we are using.

We begin by recalling the Regge-Hu expansion \cite{ReggeHu}. The Wigner $D$-functions $D^j_{\alpha' \alpha}(g(x))$ determine a basis of functions of the symmetry-group of the model. Recall that we can use group elements $g(x)$ to coordinatize the physical space that has the $S^3$ topology.  Let $q^0_{ab}(x,t)$ be the metric of an homogeneous space. A general perturbations to the three-metric $h_{ab}(x,t)=q_{ab}(x,t)-q^0_{ab}(x,t)$ can first be translated into a matrix of space-scalars $ h_{IJ}(x,t)$ by projecting it on the invariant one-forms, that is
\be \label{scalars}
h_{ab}(x,t)= h_{IJ}(x,t) \, \omega^I_a(x)\,  \omega^J_b(x);
\ee
Regge and Hu decompose these scalars in terms of definite angular-momentum components of the three-metric $h_{IJ}^{j\,\alpha}(x,t)$, labelled by spin and magnetic numbers $\{ j,\,\alpha \}$. 
\be \label{regge}
h_{IJ}(x,t)= \sum_{j,\alpha}\ h^{j\alpha}_{IJ}(x,t)\,.
\ee
These can be expressed in terms of Wigner $D$-functions $D^j_{\alpha' \alpha}(g(x))$ (whose form is recalled in Appendix \ref{WIgner}) via (see \cite{ReggeHu})
\be \label{regge2}
h^{j\alpha}_{IJ}(x,t)= \sum_{\alpha' =-j}^{j} \, h_{IJ}^{j\alpha\alpha'}(t) \, D^j_{\alpha' \alpha}(g(x));
\ee
The time dependent amplitudes $h_{IJ}^{j\alpha \alpha} (t)$ represent, at fixed $j$, $(2j+1)^2$ inhomogeneous degrees of freedom. They are governed by a set of coupled differential equations, studied in \cite{ReggeWheeler}. 

At a first sight, the dependence of the scalar harmonic function $h^{j\alpha}_{IJ}(x,t)$ on the modes $j,\alpha$ in equations (\ref{regge}-\ref{regge2}) seems to be incompatible with the Peter-Weyl decomposition \cite{Fuchs} of $h_{IJ}(x,t)$.
However, the choice of not summing over the $j,\alpha$ labels contracted with the Wigner $D$ functions relies on the evidence, due to Einstein equations, that $j, \alpha$ states can be de-coupled from $\alpha'$ states. This leaves as a possible choice to fix perturbations of the metric with definite $j, \alpha$. Indeed, only the $\alpha'$ states are mixed by the action of the derivative operators and thus by the linearized Einstein tensor, as explained in Appendix \ref{WIgner}. This can be understood moving from the action (\ref{Dvar}) of the invariant operators on the Wigner $D$ functions and expressing it in terms of derivatives vector fields on the Euclidean space $E^4$ in which $S^3$ can be embedded.

Let us adapt this formalism to the first order variables we use. 
We start with the triads. Let us write a generic perturbed triad
${E}_{I}^a(x,t)$ as the sum of the background triad $e_I$ 
field and a perturbation. 
\begin{equation} \label{inho}
{E}_{I}^a(x,t)={e}_{I}^a(x)+\psi_{I}^a(x,t)\,.
\end{equation}
It is convenient to project the perturbation on the background triad
\begin{equation} \label{inho}
\psi^a_{I}(x) = \psi_{IJ}(x,t)\ e_{J}^a(x)\,.
\end{equation}
Following Regge and Hu, we write this as a sum of
components of definite $j$ and $\alpha$ quantum numbers 
\begin{eqnarray}
\psi_{IJ}(x,t)= \sum_{ j\,\alpha}\, \psi^{ j\,\alpha}_{IJ}(x,t)\,\, , 
\end{eqnarray}
where
\begin{eqnarray}
\psi^{ j\,\alpha}_{IJ}(x,t) 
=
\,\sum_{ \alpha'=-j}^j \, \psi_{IJ}^{j\alpha\alpha'}(t) \,  D^j_{\alpha' \alpha}\left(g(x)\right)\,.
\end{eqnarray}
The same can be done for the connection 
\begin{equation} \label{inho}
\omega^{I}_a(x) \rightarrow\, \tilde{\omega}^{I}_a(x,t)=\omega^{I}_a(x)+\varphi^{I}_a(x,t)\,.
\end{equation}
Project $\varphi^I_a$ on the invariant one-forms
\begin{equation} \label{inho}
\varphi_a^{I}(x) = \varphi^{IJ}(x,t)\ \omega^{J}_a(x)\,.
\end{equation}
Expanding this in  components of definite $j$ and $\alpha$ quantum numbers gives
\begin{eqnarray}
\varphi^{IJ}(x,t)= \sum_{ j\,\alpha}\, \varphi_{ j\,\alpha}^{IJ}(x,t)\,\, , 
\end{eqnarray}
where
\begin{eqnarray}
\varphi_{ j\,\alpha}^{IJ}(x,t) 
=
\,\sum_{ \alpha'=-j}^j \, \varphi^{IJ}_{j\alpha\alpha'}(t) \,  D^j_{\alpha' \alpha}\left(g(x)\right)\,.
\end{eqnarray}
The $(\varphi^{IJ}_{j\alpha\alpha'}(t),\psi^{IJ}_{j\alpha\alpha'}(t))$ are the time-dependent expansion coefficients that capture the inhomogeneous degrees of freedom. They, are given by matrices in the internal indices $I,J$, labeled by the spin $j$ that runs from $j=1/2$ to all the semi-integers numbers, and the corresponding magnetic number $\alpha$. 

\subsection{Dipole model}

Suppose we restrict the geometry by assuming that the matrices $(\varphi^{IJ}_{j\alpha\alpha'}(t),\psi^{IJ}_{j\alpha\alpha'}(t))$ are diagonal in the internal indices $I,J$ and it is different from zero only for lowest nontrivial integer spin $j=1$ and for, say, $\alpha=0$. That is, we restrict to the components
\be
 \varphi^{IJ}_{1,0,\alpha}(t)=\delta^{IJ}\varphi^{I}_{\alpha}(t). \quad \quad
 \psi^{IJ}_{1,0, \alpha}(t)=\delta^{IJ}\psi^{I}_{\alpha}(t).\ee
where $\alpha=-1,0,1$. 
Then we have that the inhomogeneities are determined by precisely nine plus nine variables $\varphi^{I}_{\alpha}(t), \psi^{I}_{\alpha}(t)$, namely nine degrees of freedom. 
Assuming that the Gau\ss\, constraint reduces the degrees of freedom by three (see below), we obtain six degrees of freedom, which is the number of degrees of freedom captured by the dipole
variables. Thus, such a geometry is entirely captured by the six degrees of freedom of the dipole model.  Therefore, we can interpret the six extra degrees of freedom of the dipole model (beyond anisotropies), as a description of the diagonal part of the lowest integer mode of the inhomogeneities

If we do so, we can relate the variables of the dipole model to the quantities  $(\varphi^{I}_{\alpha}(t), \psi^{I}_{\alpha}(t))$. 
Notice that the one forms $\tilde\omega^I$ no longer satisfy the Maurer-Cartan equation (\ref{MC}). 
The fiducial algebra-elements $\omega_f^I$ are therefore perturbed as well. At first order, for a generic perturbation, let us define 
\ba
\tilde\omega_f^I&=&\f12\int_f\epsilon^I\,_{JK}\,\tilde\omega^J\wedge\tilde\omega^K=\omega_f^I+ \int_f\,\,\epsilon^{I}\,_{JK}\,\omega^J\wedge\varphi^K =\omega^I_f+\sum_{j\alpha\alpha'}\,\varphi^{KL}_{j \alpha\alpha'}(t) \ \phi_{f\,KL}^{I\,j\alpha\alpha'}
\ea
where
\ba
\phi_{f\,KL}^{I\,j\alpha\alpha'}=
\int_f\,\,\epsilon^{I}\,_{JK}\, \, D^j_{\alpha\, \alpha'}\,
 \,\omega^J\wedge \omega^{L}.
\ea
In particular, if we restrict to the diagonal $j=1, \alpha=0$ case,
\ba
\tilde\omega_f^I&=& \,\,\omega_f^I
+ \,\varphi^{(I)}_{\alpha}(t)\ \phi^{(I)}_{f,\alpha}
\label{questa}
\ea
where 
\ba
\phi^{I}_{f,\alpha}&=&\int_f\,\,\epsilon^{I}\,_{JK}\, \, D^{1}_{0\, \alpha}\,
 \,\omega^J\wedge \omega^{K}
 \ea
 are fixed coefficients. 
Then the relation with the dipole variables can be written as 
\be\label{varbix2}
U_f(c^I,\varphi^{I}_{\alpha})=\exp\left(c^I\tilde\omega^I_f\tau_I\right)\exp\left(\alpha\,\omega^I_f\tau_I\right), 
\ee
which replaces (\ref{varbix}).  
Similarly, we can write
\ba
\tilde E_f^I&=&\int_f (e_I^a+\psi^a_I)\epsilon_{abc}dx^b\wedge dx^c= E^I_f
+ 2  \int_f\,\,\psi_{IJ}\,\,  \epsilon^{J}\,_{KL}\,\omega^K\wedge\psi^L 
=E^I_f
+  2\sum_{j\alpha\alpha'}\, \psi_{IJ}^{j\alpha} \phi_{f, \alpha \alpha'}^J
\ea
In particular, if we restrict to the diagonal $j=1$, $\alpha=0$ case,
\ba
\tilde E_f^I&=& E^I_f + 2\,\psi_{I}^{\alpha}\  \phi_{f,\alpha}^I. 
\ea

The Gau\ss\, constraint no longer identically vanishes. It can be split into two parts: the homogeneous and the inhomogeneous terms 
\be 
\mathcal{G}^I=\sum_f p^{(I)}\omega_f^{(I)}+
2\,\psi_{I}^{\alpha}\ \sum_f \phi^{I}_{f,\alpha}\approx0\,.
\ee
The first part is the constraint which appears in (\ref{gaussbix}) within the Bianchi IX framework and vanishes identically  because of the Stokes theorem. The second gives 
three conditions on the inhomogeneous perturbations to the electric fields $\tilde E_f$. 

\section{Conclusions and perspectives}

We have studied a triangulated (loop) quantum cosmology model, and analyzed the geometry it describes.  In particular, the model of a dipole $SU(2)$-lattice theory, triangulating a topological three-sphere by means of two tetrahedra, has been related to a Bianchi IX cosmological model perturbed by six inhomogeneous degrees of freedom.  

This work is mainly based on the case of the dipole cosmology, for which the algebra of Hamiltonian constraint closes.   Higher modes of the mode expansion can be captured with finer triangulations, but these require a more work for a consistent definition.

The truncation we use relies on spherical topology. It is likely
that the technique we have used could be extended to general
compact spaces, but we have not studied this issue. For the
extension to non-compact spaces, on the other hand, the problem
is delicate. To use the same technique, one should partition a
non-compact spatial slice with ``fiducial boxes'', as is done in LQC,
but this would reduce the inhomogeneities that the model
can capture. Doing so, on the other hand, would be interesting for
the comparison with LQC.

We have only discussed the dynamics generated by the Euclidean part of the Hamiltonian constraint and with a fixed value of the Barbero-Immirzi parameter. The full Hamiltonian constraint, the possibility of using of a generic value of the Barbero-Immirzi parameter $\beta$, and the Lorentzian theory, will be considered elsewhere.

Finally, notice that the results of this paper, together with the link between
LQC and spin-foams derived in \cite{ACHVR}, might provide a path to connect cosmological spin-foam models from the cosmological sector of LQG.

\section*{Acknowledgments}

We thank Eugenio Bianchi, Martin Bojowald, Laurent Freidel and Matteo Smerlak for useful discussions. M.V.B. and A.M. gratefully acknowledge ``Fondazione Angelo Della Riccia'' and Universit\`a di Roma ``Sapienza'' for financial support during their permanence at CPT in Marseille.

\appendix 

\section{The three-sphere}  \label{tresfera}

We collect in this Appendix some useful information on the structure of a three sphere, and a few other useful formula. In \ref{ttopo} we illustrate the topology of the three-sphere and its relation to the Hopf fibration. In subsection \ref{su}, using the isomorphim of groups $S^3/\{1,-1\}\sim SO(3)$, in which the subgroup $\{1,-1\}$ is the kernel of the group homomorphism $h: S^3 \rightarrow SO(3)$, and identifying with $SU(2)$ the double cover of $SO(3)$, namely $Spin(3)$, we clarify which is the role of the group $SU(2)$ in describing the topology of $S^3$ by means of the Hopf fibration. In subsection \ref{sa}, we describe how the $SU(2)$ symmetry structure enter the definition of the Cartan one-forms. 

\subsection{Topology of the three-sphere: Hopf fibration} \label{ttopo}

We illustrate here a concrete realisation of the Hopf map \cite{Hopf} acting on $S^3$, which defines a fibration with fiber space $S^1$ and the base space $S^2$. In what follows we identify $\mathbb{R}^4$ with $\mathbb{C}^2$ and $\mathbb{R}^3$ with $\mathbb{C}\times \mathbb{R}$. Taking $(x_1,x_2,x_3,x_4)\in \mathbb{R}^4$ and $(\xi_1,\xi_2,\xi_3)\in \mathbb{R}^3$ the identification is achieved respectively by means of $z_0=x_1+i\,x_2$ and $z_1=x_3+i\,x_4$ for $\mathbb{R}^4$, and $z=\xi_1+i\xi_2$ and $y=\xi_3$ for $\mathbb{R}^3$. A unit three-sphere embedded in $\mathbb{R}^4$, as $(x_1)^2+(x_2)^2+(x_3)^2+(x_4)^2=1$, reads in terms of the complex variables $|z_0|^2+|z_1|^2=1$.  A unit two-sphere is identified with the subset of $\mathbb{C}\times \mathbb{R}$ such that $|z|^2+y^2=1$. Let us parametrize $S^2$ by means of the expression $(\xi_1)^2+(\xi_2)^2+(\xi_3)^2=1$ and take a natural bundle projection by the Hopf map $\pi\,:S^3 \to S^2$
\ba
\xi_1&=&2(x_1\,x_3+x_2\,x_4),\\\nonumber 
\xi_2&=&2(x_2\,x_3-x_1\,x_4),\\\nonumber
\xi_3&=&-(x_1)^2-(x_2)^2+(x_3)^2+(x_4)^2.
\ea
With this mapping the relation $(\xi_1)^2+(\xi_2)^2+(\xi_3)^2=\left((x_1)^2+(x_2)^2+(x_3)^2+(x_4)^2\right)^2=1$ holds.

We can now consider a stereographic projection \cite{NS, Lyons} of a point in the southern hemisphere $S^2_{-}\subset S^2$ from its north pole and label its coordinates as $U,V$. On the complex plane $\mathbb{C}$ containing the equator of $S^2\subset \mathbb{C}\times \mathbb{R}$, we consider the function $T=U+i\, V$, which is within the circle of unit radius on the complex plane and is given by
\be
T=\frac{\xi^1+ i \,\xi^2}{1- \xi^3}= \frac{x_1+ i\, x_2}{x_3+ i\, x_4}= \frac{z_0}{z_1}\,,
\ee
with $\xi^i\in S^2_{-}$. For a $\lambda\in U(1)$ such that $|\lambda|=1$, we can see that $T$ is invariant under $(z_0,\,z_1)\to (\lambda z_0,\,\lambda z_1)$, which are both points in $S^3$. Since the set of complex numbers $\lambda$ with $|\lambda|^2=1$ form the unit circle in the complex plane, it follows that for each point $\bar{m} \in S^2$, the inverse image $\pi^{-1}(\bar{m})$ is a circle, {\it i.e.} $\pi^{-1}(\bar{m}) \simeq S^1$. Thus the 3-sphere is realized as a disjoint union of these circular fibers.

Similarly, the stereographic coordinates $P,\,Q$ of the northern hemisphere $S^2_+$ projected from the south pole read 
\be
S=\frac{\xi^1- i \,\xi^2}{1+ \xi^3}= \frac{x_3+ i\, x_4}{x_1+ i\, x_2}= \frac{z_1}{z_0}
\ee
with $\xi^i\in S^2_{+}$. On the equator, $S^2_{+}\cap  S^2_{-}$, one finds $T=S^{-1}$.

The fiber bundle structure is then given by the following local trivialization\footnote{Each local trivialization is well defined on each chart, while on the equator they become $\phi_-^{-1}\,: (z_0,\,z_1)\to (z_0/z_1,\, \sqrt{2} z_1)$ and $\phi_+^{-1}\,: (z_0,\,z_1)\to (z_1/z_0,\, \sqrt{2} z_0)$: the transition function on the equator is $\tau_{+-}(\xi)=(\sqrt{2}z_0)/(\sqrt{2}z_1)=\xi_1 + i\, \xi_2 \in U(1)$. While following a path around the equator, the transition function $\tau_{+-}(\xi)$ crosses the unit circle in the complex plane once. Thus the $U(1)$ bundle $S^3 \to^{\!\!\!\!\!\!\pi} \,\,S^2$ is characterized by the homotopy class $1$ of $\pi_1(U(1))=\mathbb{Z}$. The Hopf fibration is therefore locally a product space, but it is not a trivial fiber bundle, as $S^3$ is not (globally) a product of $S^2$ and $S^1$.}: on the south hemisphere one defines $\phi_-^{-1}\,:\pi^{-1}(S^2_-) \to S^2_- \times U(1)$ as given by $(z_0,\,z_1)\to ( z_0/z_1,\, z_1/|z_1|)$ while on the north hemisphere the quantity $\phi_+^{-1}\,:\pi^{-1}(S^2_+) \to S^2_+ \times U(1)$ is given by $(z_0,\,z_1)\to ( z_1/z_0,\, z_0/|z_0|)$. This construction clarifies that $S^2$ is the base space of $S^3$.

\subsection{$SU(2)$ description of the three-sphere} \label{su}

A further geometric interpretation of the Hopf fibration can be obtained considering rotations of the two-sphere in a three-dimensional space \cite{Lyons}. We recall that ${\it Spin}(3)$ is the double cover of the rotation group $SO(3)$ and that is diffeomorphic to $S^3$. The spin group acts in a transitive way on $S^2$ by rotations. The stabilizer subgroup\footnote{For every element $x$ of a set $X$, the stabilizer subgroup of $x$ is the set of all elements in $G$ that fix $x$.} of a point of $S^2$ is isomorphic to the circle group, from which it follows that $S^3$ is a principle circle bundle over $S^2$. This is exactly the Hopf fibration. 

Concretely, this can be seen identifying ${\it Spin}(3)$ with the group of unit quaternions ${\it Sp}(1)$: a point $(x_1,\,x_2,\,x_3,\,x_4)\,\in\, \mathbb{R}^4$ is interpreted \cite{Kuipiers} as a quaternion $q\in\mathbb{H}$ by $q=x_1+{\bf i} \,x_2+{\bf j}\,x_3+{\bf k}\,x_4$ and the three-sphere is then identified with the quaternions of unit norm, namely $q\,\bar{q}=|q|^2=(x_1)^2+(x_2)^2+(x_3)^2+(x_4)^2=1$. Now a vector $p$ in $\mathbb{R}^3$ can be identified with the imaginary part of a quaternion, {\it i.e.} $p={\bf i}\,p_1+{\bf j}\,p_2+{\bf k}\,p_3$, and the mapping $p \to q\,p\,\bar{q}$ is a rotation in $\mathbb{R}^3$, and thus an isometry as $|q p\bar{q}|^2=|p|^2$. Morover it is not hard to check that this mapping, as a rotation, preserves orientation. The $q$ are then unit quaternions provided by the group of rotation in $\mathbb{R}^3$, with opposite elements $-q$ undergoing the same transformations. The set of unit quaternions $q$ which fixes a unit imaginary quaternion have the form $q=\zeta + \nu\, p$, with $\zeta, \, \nu \in \mathbb{R}$ such that $|\zeta|^2+|\nu|^2=1$ ({\it i.e.} a circle subgroup). Taking then $p={\bf k}$, one can define the Hopf fibration via the map $\pi\,: q \to q {\bf k} \bar{q}$. The image set by the Hopf map is made of (still unit) quaternions provided only by imaginary parts: these points lie on the two-sphere $S^2$. As a quaternion $q=x_1+{\bf i} \,x_2+{\bf j}\,x_3+{\bf k}\,x_4$ can be recast in terms of a $2\times 2$ matrix, 
\be \label{qsu2}
q (x_1,\,...\,x_4) := \begin{pmatrix}  x_1 +{\bf i} \, x_2  & x_3 + {\bf i} \, x_4\\
                                -x_3 +{\bf i} \, x_4  & x_1 - {\bf i} \, x_2
        \end{pmatrix}\,,      
\ee
one can identify the group of unit quaternions with $SU(2)$ and represent imaginary quaternions by the skew-hermitian $2\times 2$ matrices, which are isomorphic to $\mathbb{C}\times \mathbb{R}$. 
The fiber for a given point of $S^2$ consists now of all those unit quaternions whose image via the Hopf map $\pi\,: q \to q {\bf k} \bar{q}$ aims there. These points are easily recognised to belong to a circle. A direct way of seeing it is to consider that the multiplication by unit quaternions is equivalent to realise a composition of rotations in $\mathbb{R}^3$: for instance multiplying by $q_t= e^{{\bf k} t} $ corresponds to a rotation by $2t$ around the $z$ axis --- varying $t$ one draws a great circle of $S^3$. As long as a generic base point $\beta=(l,m,n)$ is not the antipode $(0,0,-1)$, any quaternion $q_{\beta}$ (associated to the base point $\beta$ by means of $\pi^{-1}$) not having component on the $z$ axis, say for instance $q_{\beta}=(1+n, -m,l,0)/\sqrt{2(1+n)}$, will continue aim there after multiplication by $q_t$: the fiber of the base point $\beta$ is therefore given by quaternions of the form $q_{\beta}\cdot q_t$, which does not represent merely a topological circle, but a geometric one \cite{Berger}. For the base point $(0,0,-1)$, whose associated quaternion can be taken to be the ${\bf i}$ axis, the fiber is simply $(0, \cos t, -\sin t, 0)$.

\subsection{Differential structure of the three-sphere} \label{sa}

The unit quaternions, once expressed by means of the unimodular (because of the $q \bar{q}=1$ condition identifying a $S^3$) matrices (\ref{qsu2}), can be in turn represented by means of {\it Euler angles}, which are another set of coordinates $\{ \theta,\, \phi,\,\psi \}$ on the three-sphere such that $\theta$ and $\phi$, as usual, are in the range $\theta \in [0, \pi)$, $\phi \in [0, 2 \pi)$, while for the Euler angle $\psi$ one must set $\psi\in [0, 4 \pi)$ in order to achieve a simply connected covering\footnote{This chart breaks down at the poles $\theta=\phi=\psi=0$, corresponding to the identity $1\!\!1$, and at $\theta=\phi=0$ and $\psi=2\pi$, corresponding to $- 1\!\!1$.} space with the topology of $S^3$.
By these one $q$ assumes the form
\be\label{qeu}
q(\theta,\, \phi,\, \psi):= e^{-\phi \tau_1 } \, e^{-\theta \tau_2 } \, e^{-\psi \tau_3 }\,.
\ee
In terms of Euler angles, isometries of the three-sphere are well understood as left (and right) translation generated by $SU(2)$ representations. Recall that a left (right) translation is a mapping $\mu: S^3 \to S^3$ which take a point ${\bf P}$ in a point  point ${\bf P'}$, where $q({\bf P})= \mu\,q({\bf P'}) $ (and respectively $q({\bf P})= q({\bf P'})\, \mu $). These are really translations, as the only mapping having fixed points is the identity. Otherwise, a mapping having fixed points can be defined by $q({\bf P'})= \mu ({\bf P}) \mu^{-1}$. This latter is clearly a rotation.

The right and the left translation can be used to define differential operators on $S^3$. For instance a right translation can be represented by a hermitian matrix $\upsilon$ and a parameter $v$, for which $q({\bf P'})= q({\bf P})\, e^{i\, v \upsilon}$ or by a differential operator $\mathcal{K}$ and a parameter $v$, such that  $q({\bf P'})= [e^{iv \mathcal{K}}q]({\bf P})$. The compatibility condition for those two representation is $q\,\upsilon = \mathcal{K} q$.

Similarly, a left translation is defined by $\upsilon \, q = \tilde{\mathcal{K}} q$, where using (\ref{qsu2}) and previous definition of left and right translation, it is possible to see that $\tilde{\mathcal{K}}(x_1,\,x_2,\,x_3,\,x_4)=\bar{\mathcal{K}}(-x_1,\,-x_2,\,-x_3,\,x_4)$. Similarly, in term of Euler angles, one must send $(\phi,\,\theta,\,\psi)\to(-\psi,\,-\theta,\,-\phi)$. Right and left translation operators can then be determined from these definitions. If we relabel a point in $\mathbb{R}^4$ with $(x_1,\,x_2,\,x_3,\,x_4\,) =\, (x_i,X_4)$, we can recognize, respectively, the right and the left translation operators to be \cite{MT}
\ba
J_k &=& \frac{i}{2} \left( x^k \frac{\partial}{\partial x_4 } -x_4 \frac{\partial}{\partial  x^k}  - x^s \epsilon_{skl}  \frac{\partial}{\partial  x^l} \right) \\ \nonumber 
\tilde{J}_k &=& \frac{i}{2}  \left( x^k \frac{\partial}{\partial x_4 } -x_4 \frac{\partial}{\partial  x^k}  + x^s \epsilon_{skl}  \frac{\partial}{\partial  x^l} \right) 
\ea
Notice that these generators\footnote{Rotations about the fixed point $x_1=x_2=x_3=0$ and $x_4=1$ are generated by the operators $L_k=J_k-\tilde{J}_k$.} fulfill two $su(2)$ algebras
\ba
[J_i,\, J_j]= i\, \epsilon_{ijk} J_k\,, \qquad [\tilde{J}_i,\, \tilde{J}_j]= i\, \epsilon_{ijk} \tilde{J}_k\,, \qquad [\tilde{J}_i,\,J_k]=0\,.
\ea
These four-dimensional operators are all tangent to the surfaces of constant modulus, represented by $(x_1)^2+(x_2)^2+(x_3)^2+(x_4)^2=e^t$, with $t\in \mathbb{R}$. Thus, focusing on our case $t=0$ and equating relation (\ref{qsu2}), which is written in cartesian coordinates, to relation (\ref{qeu}) expressed in terms of Euler angles, one finds change of variables and then translation operators in terms of Euler angles.

Making a further transformation, which defines Hopf coordinates in terms of Euler angles,
\be \label{euho}
\eta=\frac{1}{2}\theta\,, \qquad \xi_+=\frac{1}{2}(\psi+\phi)\,, \qquad \xi_-= \frac{1}{2}(\psi-\phi)\,,
\ee
one finds the transformation from cartesian to Hopf coordinates 
\ba \label{caho}
x_1= \sin(\theta/2) \, \sin \xi_-\,, \qquad x_2= \sin(\theta/2) \, \cos \xi_- \,, \qquad x_3= \cos(\theta/2) \, \sin \xi_+\,, \qquad x_4= \cos(\theta/2) \, \cos \xi_+\,.
\ea 
Transformations (\ref{euho}) and (\ref{caho}) allow us to think at the topology of $S^3$ in terms of two-torus coordinates, as both $\xi_1,\,\xi_2,\,\in[0,\,2\pi)$. 
In fact, we can consider the third component of the left and right invariant vector fields $J_3$ and $\tilde{J}_3$, and their combinations $J_3-\tilde{J}_3$ and $J_3+\tilde{J}_3$. Their duals one-forms are now linear in $d\xi_-$ and  $d\xi_+$, as
\ba
\omega^3-\tilde{\omega}^3&=& -2 \cos^2 \eta \,d\xi_+\,, \\\nonumber
\omega^3+\tilde{\omega}^3&=& -2 \sin^2 \eta \,d\xi_- \,.
\ea
The metric on the unitary $3$-sphere turns out to be expressed by
\be
dl^2= 4\left( (d\eta)^2 + \sin^2 \eta \,(d\xi_-)^2+ \cos^2 \eta \,(d\xi_+)^2  \right)\,,
\ee
which makes clear that the trajectories of these two-parameter subgroup are constant $\eta$ surfaces and that they shrink in the $\xi_-$ direction near the degenerate trajectory $\eta=0$, and in the $\xi_+$ direction near the other degenerate direction $\eta= \pi/2$. In terms of the Hopf fibration, each torus, which topologically is in stead the product of two circles, can be thought as the stereographic projection of the inverse image of a circle of latitude of the two-sphere. 

\section{Traces of $\tau$-matrices}\label{tau}

We give here results of straightforward calculations of traces of products of $\tau$-matrices. We remind that  a basis for the $su(2)$ algebra elements  in the fundamental representation is given by $\tau_I=\sigma_I/(2i)$, in which $\sigma_I$ are intended to be the Pauli matrices. It follows that:
\ba \nonumber
\tr[\tau_I\tau_J]&=&-\f12\delta_{IJ}\,,\\\nonumber
\tr[\tau_I\tau_J\tau_K]&=&-\f14\epsilon_{IJK}\,,\\\nonumber
\tr[\tau_I\tau_J\tau_K\tau_L]&=&-\f14\left(\delta_{IK}\delta_{JL}-\delta_{IL}\delta_{JK}\right)+\f18\delta_{IJ}\delta_{KL}\,,\\\nonumber
\tr[\tau_I\tau_J\tau_K\tau_L\tau_S]&=&\f14\left(-\f14\delta_{KS}\epsilon_{IJL}+\f14\delta_{KL}\epsilon_{IJS}+\delta_{LS}\epsilon_{IJK}+\delta_{IJ}\epsilon_{LSK}\right)\,, \\ \nonumber
\tr[\tau_I\tau_J\tau_K\tau_L\tau_S\tau_M]&=&\f18\left(-\f14\epsilon_{IKL}\epsilon_{JSM}+\f14\epsilon_{JKL}\epsilon_{ISM}+\f12\delta_{SM}(\delta_{IK}\delta_{JL}-\delta_{IL}\delta_{JK})+\right.\\
&+&\left.\f12\delta_{KL}(\delta_{IS}\delta_{JM}-\delta_{IM}\delta_{JS})+\f12\delta_{IJ}(\delta_{KS}\delta_{LM}-\delta_{KM}\delta_{LS})-2\delta_{IJ}\delta_{KL}\delta_{SM}\right)\,.
\ea

\section{Wigner $D$-functions}\label{WIgner}

Finally, we recall here an explicit formula for the Wigner $D$-functions. 
These are obtained from their very definition in terms of matrix elements of the rotation operator $\mathcal{R}(\phi,\theta,\psi)=\exp(-i \phi \bf{j}_x)\,\exp(-i \theta \bf{j}_y)\,\exp(-i \psi \bf{j}_z)\,$, in which  $\bf{j}_x\,,\bf{j}_y, \bf{j}_z$ are generators of the $su(2)$ Lie-algebra, and can be expressed as \cite{Var}
\ba
&D^j_{\alpha' \alpha}\equiv \langle j\, \alpha' | \mathcal{R}(\phi,\theta,\psi) | j\, \alpha \rangle= e^{-i \alpha' \phi} \, d^j_{\alpha' \alpha}(\theta) \,e^{-i \alpha \psi} \,,   \nonumber\\
&\!\!\!\!\!\!\!\!\!d^j_{\alpha' \alpha}(\theta)= [(j+\alpha')!(j-\alpha')!(j+\alpha)!(j-\alpha)!]^{1/2}
\sum_s \frac{(-1)^{\alpha'-\alpha+s}}{(j+\alpha-s)!s!(\alpha'-\alpha+s)!(j-\alpha'-s)!}  \left(\cos\frac{\theta}{2}\right)^{2j+\alpha-\alpha'-2s}\left(\sin\frac{\theta}{2}\right)^{\alpha'-\alpha+2s}\!\!\!,
\ea
in which the sum over $s$ is over such values that the factorials are nonnegative.

For $j=l$ integers, the Wigner $D$-functions are simply related to the spherical harmonics $Y_{l,\alpha}(\theta,\phi)$ by the relation 
\be
D^l_{\alpha' \alpha}(g)= (-1)^{-\alpha'} \sqrt{4 \pi/(2l+1)} \,Y_{l, -\alpha'}(\theta, \phi)\, e^{i \alpha\, \psi}\,.
\ee

More in particular, the Wigner $D$-functions, which are the representation functions of the $SO(3)$ and $SU(2)$ simmetry-groups, can be obtained 
by imposing that they satisfy the following differential equations, expressed in terms of the Euler angles
\ba
&\hat C\, D^j_{\alpha' \alpha}=
\left[   
\frac{\partial^2}{\partial \theta^2}+\cot \theta\, \frac{\partial}{\partial \theta}+ \frac{1}{\sin^2 \theta}
 \left( \frac{\partial^2}{\partial \phi^2} - 2 \cos \theta\, \frac{\partial^2}{\partial \phi\,\partial \psi} + \frac{\partial^2}{\partial \psi^2} \right)
\right]
\, D^j_{\alpha' \alpha}= j(j+1)\,D^j_{\alpha' \alpha}\,, \nonumber\\
&\hat L_3\, D^j_{\alpha' \alpha}=-i\frac{\partial}{\partial \psi}\,D^j_{\alpha' \alpha} =\alpha'\,D^j_{\alpha' \alpha}\,, \qquad
\hat L_z\, D^j_{\alpha' \alpha}=-i\frac{\partial}{\partial \phi}\,D^j_{\alpha' \alpha} = \alpha\,D^j_{\alpha' \alpha}\,. \qquad
\ea

In the above equation we have introduced two bases of generators for the $su(2)$ algebra and their common Casimir operator $\hat C$, which is an invariant of the group. The angular momentum operators $\{\hat L_1,\,\hat L_2,\,\hat L_3\}$ of the three-dimensional rotation group in quantum mechanics, which are the intrinsic angular momentum operators of a rigid body, are related to the left-invariant vector fields $e_I^a$ (which generate right-transformations) via the relations 
\be \label{mela}
\hat L_1= i\, e_1^a\, \partial_a\,, \qquad 
\hat L_2= i\, e_2^a\, \partial_a\,, \qquad 
\hat L_3= i\, e_3^a\, \partial_a\,. 
\ee
The spatial angular momentum operators $\{\hat L_x,\,\hat L_y,\,\hat L_z\}$ are in turns related to the right-invariant Killing vectors $\xi_I^a$ (which generates left-transformations) by the formulae
\be
\hat L_{x}=-i \xi_1^a\, \partial_a\,, \qquad
\hat L_{y}=-i \xi_2^a\, \partial_a\,, \qquad
\hat L_{z}=-i \xi_3^a\, \partial_a\,. \qquad
\ee
In the chart provided by the Euler angles, it is straightforward to find the expression for the left invarint frames $e_I$ as well as that one for the right invariant vector fields $\xi_I$ (which Lie drag the the frame and coframe introduced in Section \ref{Bianchi9}). For the frames 
\ba
e_{1} &=& -\cos\psi \f{\partial}{\partial\theta} - \f{\sin\psi}{\sin\theta}\f{\partial}{\partial\phi} +\f{\cos\theta\sin\psi}{\sin\theta}\, \f{\partial}{\partial \psi}\,, \nonumber\\
e_{2} &=& \sin\psi \f{\partial}{\partial\theta} - \f{\cos\psi}{\cos\theta}\f{\partial}{\partial\phi} + \f{\cos\theta \cos\psi}{\sin\theta} \f{\partial}{\partial \psi}\nonumber\,,\\
e_{3} &=& \f{\partial}{\partial\psi},\
\ea
which fulfill the relation $[e_I,e_J]=-\epsilon_{IJ}^{\,\,\,\,\,\,\,K}\,e_K$.

It turns out that the Killing vectors are in the following relation to the frames
\ba
\xi_{1} &=& \cos\phi \f{\partial}{\partial\theta} - \f{\cos\theta\cos\phi}{\sin\theta}\f{\partial}{\partial\phi} +\f{\sin\phi}{\sin\theta}\, \f{\partial}{\partial\psi} = \nonumber\\
&=& -\left(  \cos \phi \cos\psi -\cos\theta\sin\phi\sin\psi  \right)e_{1}  + \left( \cos\phi\sin\psi
+\cos\theta\sin\phi\cos\psi \right) e_{2} + \left(  \sin\phi\sin\theta  \right) e_{3} \,, \nonumber\\
\xi_{2} &=& \sin\phi \f{\partial}{\partial\theta} + \f{\cos\theta\cos\phi}{\sin\theta}
\f{\partial}{\partial\phi} - \f{\cos\phi }{\sin\psi} \f{\partial}{\partial\psi}=\nonumber\\
&=& -\left(  \cos \psi \sin\phi +\cos\theta\cos\phi\sin\psi  \right) e_{1}  + \left(  \sin\phi\sin\psi
-\cos\theta\cos\phi\cos\psi  \right) e_{2} - \left(  \cos\phi\sin\theta  \right) e_{3} \,, \nonumber\\
\xi_{3} &=& - \f{\partial}{\partial\phi} \,= \, \sin\theta\sin\psi\, e_{1} + \sin\theta\cos\psi\, e_{2} -\cos\theta \, e_{3}\,, 
\ea 
from which the Lie brackets $[\xi_I,\xi_J]=\epsilon_{IJ}^{\,\,\,\,\,\,\,K}\,\xi_K$ easily follow. Moreover 
\be
[\,\xi_I,\,e_J\,]=  \xi_I\triangleright e_J=\xi_I \triangleleft e_J  = 0\,,
\ee
which means that $e_I$ are left-invariant vector fields ({\it i.e.} they are invariant under the left action ``$\,\triangleright$'' of the Killing vectors) and, in turns, that $\xi_I$ are right-invariant vector fields ({\it i.e.} they are invariant under the right action ``$\triangleleft\,$'' of the frames).

We want now to show, following \cite{Hu}, that the action of the derivatives, expressed in terms of the action of the invariant operators, mixes only the $\alpha'$ states. That is what we briefly mentioned in Section \ref{inhomo} while we were expanding inhomogeneous perturbations (intended as scalar harmonic functions on $S^3$) as a linear combination of Wigner $D$-functions.  
 
From the previous equations it follows that
\ba \label{Dvar}
\hat L_+\, D^j_{\alpha' \alpha}&=& \left(\hat L_1 + i\, \hat L_2  \right)\,D^j_{\alpha' \alpha}= i \sqrt{(j+\alpha')(j-\alpha'+1)}\,D^j_{(\alpha'-1) \alpha}\,,\nonumber\\
\hat L_-\, D^j_{\alpha' \alpha}&=& \left(\hat L_1 - i\, \hat L_2  \right)\,D^j_{\alpha' \alpha}= i \sqrt{(j+\alpha'+1)(j-\alpha')}\,D^j_{(\alpha'+1) \alpha} \,,\nonumber\\
\hat L_3\, D^j_{\alpha' \alpha}&=& \alpha' \,D^j_{\alpha' \alpha}\,.
\ea
The invariant operators can be rewritten in terms of Cartesian coordinates $x_A=\{x_1,\,x_2,\,x_3,\,x_4\}$ in the Euclidean space $E^4$ in which $S^3$ is embedded (see Appendix \ref{tresfera}). In stead of the three co-frames $\omega^I$ in the Euler angles chart, in the Euclidean space the invariant basis is given by the four forms $\sigma^A$, which are related to $\omega^I$ by the transformation matrices $S_{IA}(x_A)$ via
\be
\omega^I=2\,S_{IA}(x_A)\, dx^A\,.
\ee

Conversely, the coordinates differentials of $E^4$ are expressible in terms of $\omega^I$ via $dx^A=1/2\,S_{IA} \omega^I$.

The coordinates derivatives, which are the vector fields of $E^4$, are then expressed by
\be
\frac{\partial}{\partial x_A}=2\,S_{IA}(x_A) \, e_I\,.
\ee

It turns out to be much easier, using the homogeneity of the $S^3$ spatial slices,  evaluate the Cartesian derivatives at the pole $x_{P}=(x_4=1,\,x_1=x_2=x_3=0)$, where  the transformations matrices reduce to $S_{IA}=-\delta_{IA}$ thus yielding the relations
\be \label{polo}
\frac{\partial}{\partial x_A}\Big|_{x_P}=-2\, e_I\,. 
\ee
Equation (\ref{polo}) allows us, by means of (\ref{mela}), to express derivatives in terms of invariant operators. On other hand, equations (\ref{Dvar}) specify the actions of the invariant operators on the Wigner $D$-functions. It follows, once the Einstein equations for the Bianchi IX model have been rewritten in terms of Cartesian coordinates $x^A$ \cite{ReggeHu}, that the perturbations to homogeneity possess states with definite $j,\alpha$ and that only perturbations labelled by $\alpha'$ states are mixed. This result has been used in Section \ref{inhomo} in order to de-couple $j,\alpha$ modes from $\alpha'$ modes.


\end{document}